\documentclass[%
superscriptaddress,
 amsmath,amssymb,
 aps,
floatfix,
]{revtex4-2}
\usepackage{graphicx}

\usepackage{algorithm}%
\usepackage{algorithmicx}%
\usepackage{algpseudocode}%
\usepackage{listings}%
\usepackage{setspace}
\usepackage{subcaption}
\usepackage{amsmath}
\usepackage{amssymb}
\usepackage{amsmath}
\usepackage{mathtools}
\usepackage{xcolor}
\usepackage[makeroom]{cancel}
\usepackage{lineno}
\usepackage{float}
\usepackage{tikz}
\usepackage{enumitem}

\numberwithin{equation}{section}
\begin{document}

\title{Intrinsic noise in structured replicator dynamics modelling time delays}
\author{Jacek Mi\c{e}kisz}\thanks{miekisz@mimuw.edu.pl}
\affiliation{Institute of Applied Mathematics and Mechanics, University of Warsaw, Banacha 2, 02-097 Warsaw, Poland}
\author{Javad Mohamadichamgavi}\thanks{jmohamadi@mimuw.edu.pl}
\affiliation{Institute of Applied Mathematics and Mechanics, University of Warsaw, Banacha 2, 02-097 Warsaw, Poland}


\begin{abstract}
We construct and analyze structured replicator dynamics of the Snowdrift game. In our model, the offspring is put in juvenile compartments and then mature and join adult compartments with strategy-dependent rates. This is augmented by death rates and hence the population size is bounded. In the corresponding birth-death Markov jump process, rates of leaving juvenile compartments may be interpreted as inverses of averages of exponentially distributed time delays. We observe a novel behavior: for equal average time delays of both strategies, the frequency of cooperators in the quasi-stationary state of a stochastic dynamics is bigger than that in the corresponding stationary state of the deterministic structured replicator dynamics which is actually equal to the critical point of the original replicator equation for the Snowdrift game. In short, time delays favor cooperation in the presence of intrinsic stochastic fluctuations.\\[7pt]

{\bf{Keywords}}: evolutionary game theory, structured replicator dynamics, intrinsic noise, random time delays, Markov jump processes
\end{abstract}


\maketitle

\section{Introduction}\label{sec1}

The standard approach to model time evolution of populations of players is to construct appropriate replicator equations for frequencies of strategies \cite{tayloryonker,hofbauerschuster,hofbauersigmund,weibull}. One assumes that populations are infinite and well-mixed. However, real populations are finite, their growth is suppressed by an environmental carrying capacity. We would like to refer here to a paper of Tao and Cressman \cite{taocressman} which was an inspiration for our work presented here. They introduced a background fitness into replicator equations as a decreasing function of the population size, corresponding to the carrying capacity.

It is usually also assumed that interactions between individuals take place instantaneously and their effects are immediate. In reality, all processes take a certain amount of time. The effects of time delays on replicator dynamics were discussed in \cite{taowang,jalboszta}.

It is well known that time delays may cause oscillations in dynamical systems \cite{gyoriladas,gopalsamy,kuang,erneux}. 
Recently however there were constructed models with strategy-dependent time delays with a novel behavior, namely the continuous dependence of equilibria on time delays \cite{jamarekraffi,jamarek,smalldelays}. 

In all models, incorporation of time delays led to infinite-dimensional time-delayed differential equations. In a very recent paper \cite{fic}, a different approach was presented. The authors introduced a compartment for juveniles, called kindergarten, who cannot participate in games, and for adults, with appropriate rates of maturing and leaving a kindergarten to join an adult population. 

Here we combine constructions of \cite{taocressman} and \cite{fic}. The compartment model is augmented by death rates and hence the population size is bounded. 

We numerically solve our structured replicator equations for the Snowdrift game to find stationary states both for population sizes and frequencies of strategies. We show that strategy-dependent transition rates for leaving kindergartens cause continuous shifts of stationary states. We see, as in previous models, that delays are not beneficial - corresponding strategies have smaller basins of attraction. 

Then we follow ideas of Tao and Cressman \cite{taocressman} to investigate effects of intrinsic noise. We introduce a birth and death Markov jump process corresponding to the deterministic dynamics. In such a model, an intensity of leaving a juvenile compartment may be seen as an inverse of an average of an exponentially distributed time delay.

We perform stochastic simulations to get expected values of population sizes and frequencies of strategies in the quasi-stationary state.

We report a novel behavior. For equal or almost equal average time delays, we observe that time delays are beneficial for cooperation - bigger time delay, bigger the frequency of cooperators in the quasi-stationary state of the stochastic dynamics. This is a joint effect of both time delays and stochasticity.

\section{Structured replicator dynamics}\label{sec2}

In classical replicator dynamics one do not take into account evolution of the size of a population 
\cite{tayloryonker,hofbauerschuster,hofbauersigmund,weibull}. In fact it grows to infinity which is obviously not realistic. 

Tao and Cressman \cite{taocressman} introduced a background fitness into replicator equations as a decreasing function of the population size, corresponding to an environmental carrying capacity. We extend their model in order to take into account time delays which naturally appear in every process describing evolution of systems of interacting objects, such as substrates in chemical reactions or individuals in evolutionary games.
Instead of dealing with explicit time delays which leads to infinite-dimensional, not easy to analyzed, time-delayed replicator equations, as it is usually done, following \cite{fic} we introduce age compartments. Namely, the offspring is put in juvenile compartments (kindergartens) then mature and join adult compartments with strategy-dependent rates. Let us emphasize here that we do not have an explicit age structure and explicit time delays. In the corresponding birth-death Markov jump process, inverses of rates of leaving juvenile compartments may be interpreted as averages of exponentially distributed time delays.

We consider here the Snowdrift game, a two-player game with two pure strategies, cooperation (C) and defection (D), and with the following payoff matrix,
\vspace{2mm}

\begin{center}
\begin{tabular}{l|ll}
  & C & D \\ \hline
C & $b-\frac{c}{2}$ & $b-c$ \\
D & $b$ & $0$
\end{tabular}
\end{center}

\vspace{3mm}
where $b$ is a reward of coming home and $c$ is a cost of snow removal, $b>c$. 

The game has a mixed Nash equilibrium, a globally asymptotically stable interior point in the replicator dynamics. We fix here $b=6$ and $c=4$, the mixed Nash equilibrium is then equal to $0.5$. In the standard replicator dynamics, individuals are matched randomly into pairs, play games and receive payoffs which are interpreted as the number of offspring who inherit strategies of parents. It is assumed that offspring immediately join the population and can participate in games. 

To take into account time delays, we introduce a two-stage life cycle for individuals. Specifically, juveniles born from adult parents first enter a separate juvenile compartment, called kindergarten, where they must wait for a certain random time before maturing and leaving it. Then they can engage in the game and reproduce. We construct a model with two compartments: for adults (A) and for juveniles (K).

We denote the number of individuals in each compartment as $n_C^A$ and $n_D^A$ for adults, and $n_C^K$ and $n_D^K$ for juveniles $C$ and $D$ players.
The total number of individuals in each compartment is given by $N^A = n_C^A + n_D^A$ for adults and $N^K = n_C^K + n_D^K$ for juveniles. Since only adults can participate in games, the expected payoff of each strategy is given by $\pi_C = \frac{n_C^A(b-\frac{c}{2}) + n_D^A(b-c)}{N^A}$ and $\pi_D = \frac{n_C^Ab
}{N^A}$. The rate at which $i$
players interact and consequently make the corresponding juvenile compartment to grow is given by $n_i^A \pi_i$. To follow ideas in \cite{taocressman} and to take into account population size effects, we assume that players share, independent of strategies the same background fitness, $-\beta^j N^j$ with a positive parameter $\beta^j$, $j=A,K$. New juveniles enter the juvenile compartment and wait for an average period $\tau_i$, $i \in {C, D}$, to mature and join the adult population. Thus, the maturation rate is $\frac{n_i^K}{\tau_i}$. This leads to the following differential equations,

\begin{align}
    &\frac{dn_C^A}{dt}=\frac{n_C^K}{\tau_C}-\beta^AN^An_C^A\nonumber\\
    &\frac{dn_D^A}{dt}=\frac{n_D^K}{\tau_D}-\beta^AN^An_D^A\nonumber\\
    &\frac{dn_C^K}{dt}=n_C^A\pi_C-\beta^KN^Kn_C^K-\frac{n_C^K}{\tau_C}\nonumber\\
    &\frac{dn_D^K}{dt}=n_D^A\pi_D-\beta^KN^Kn_D^K-\frac{n_D^K}{\tau_D}
    \label{determinestic1}
\end{align}

Let $x = \frac{n_C^A}{N^A}$ and $y = \frac{n_C^K}{N^K}$ denote frequencies of the strategy $C$ in the adult and juvenile compartments. We can derive in the standard way the system of equations for frequencies and total populations,

\begin{figure}
    \centering
    \includegraphics[scale=0.15]{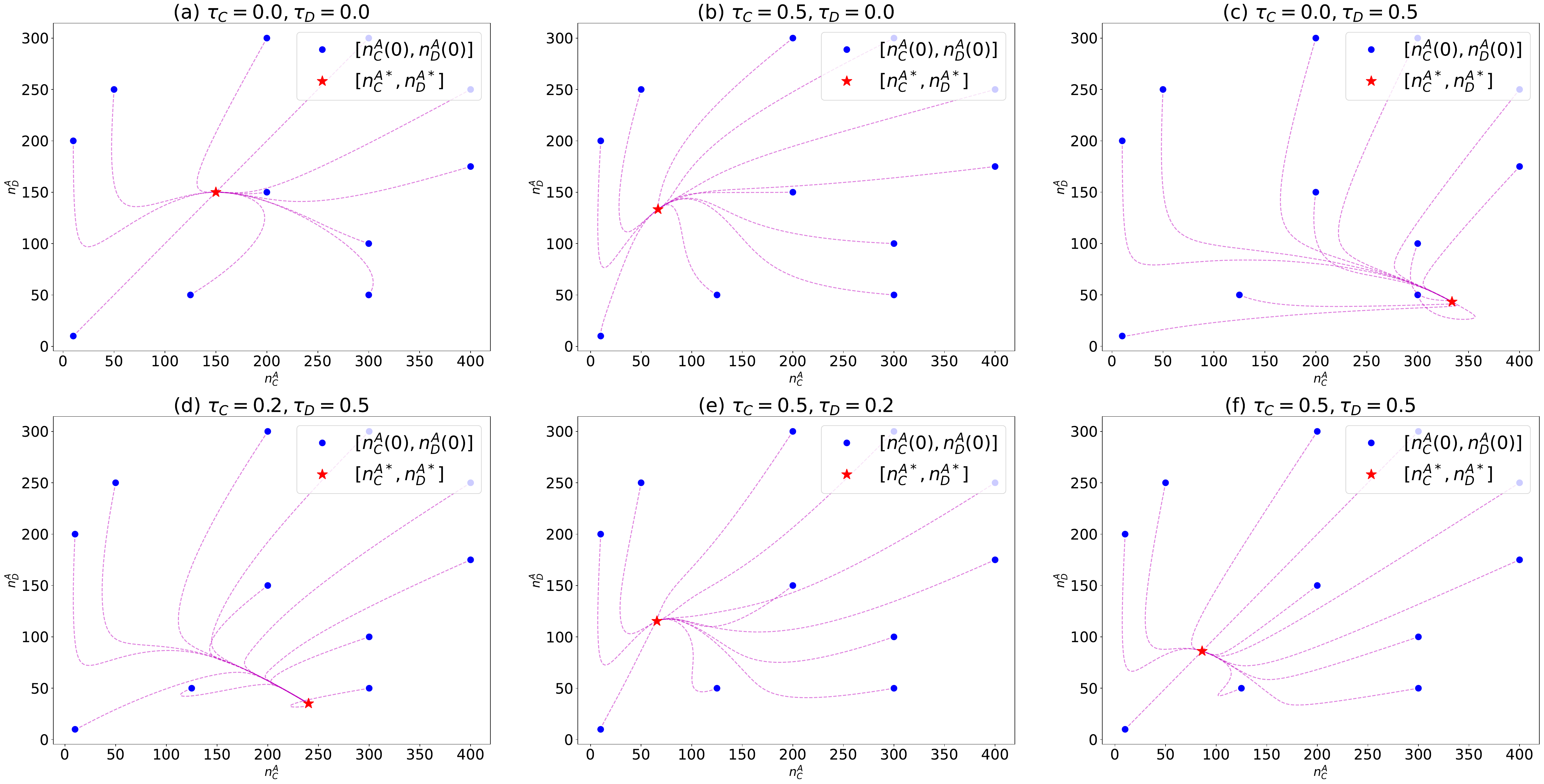}
    \caption{Phase portraits illustrating the dynamics of $(n_C^A, n_D^A)$, $\beta^A$ = $\beta^K$ = $0.01$, we solve Eq. (1) with 0 initial conditions in kindergartens. Interior stationary states $(n_C^{A*}, n_D^{A*})$ are approximately equal to a:(150,150), b:(67,133), c:(334,43), d:(240,35), e:(66,115), f:(86,86).}
    \label{fig:phaseportrate}
\end{figure}

\begin{align}
    &\frac{dx}{dt}=\frac{N^K}{N^A}\big(\frac{y}{\tau_C}(1-x)-\frac{x(1-y)}{\tau_D}\big)\nonumber\\
    &\frac{dy}{dt}=\frac{N^A}{N^K}\big(x\pi_C-y\bar{\pi}\big)+y\big(\frac{y-1}{\tau_C}+\frac{1-y}{\tau_D}\big)\nonumber\\
    &\frac{dN^A}{dt}=N^K\big(\frac{y}{\tau_C}+\frac{1-y}{\tau_D}\big)-\beta^A(N^A)^2\nonumber\\
    &\frac{dN^K}{dt}=N^A\bar{\pi}-N^K\big(\frac{y}{\tau_C}+\frac{1-y}{\tau_D}+\beta^KN^K\big)
    \label{deterministic2}
\end{align}

where $\bar{\pi}=x\pi_C+(1-x)\pi_D$. Analogous equations for cases where one time delay is equal to zero, so the corresponding rate of leaving the juvenile compartment is infinite, are presented in the Appendix A. 

For equal average delays for both strategies, 
$\tau_{C} = \tau_{D} = \tau$, our replicator equations are particularly simple and we get that stationary values of frequencies of strategies are equal to those of classical replicator dynamics without time delays. In particular, for the Snowdrift game with  $b=6$ and $c=4$, $x^* = y^* = 0.5$. Details of calculations are given in the Appendix A. 

For strategy-dependent time delays, our structured replicator equations are highly nonlinear, this makes finding stationary states analytically impossible. Below we present numerical solutions of structured replicator dynamics and stationary states of population sizes and strategy frequencies $\{n_C^{A*}, n_D^{A*}, n_C^{K*}, n_D^{K*}, N^{A*}, N^{K*}, x^*, y^*\}$.

Fig. \ref{fig:phaseportrate} presents phase portraits illustrating the dynamics of  $n_C^A$  and  $n_D^A$  for various average time delays. Here and in all following figures, the parameters  $\beta^A$ and $\beta^K$ are set to $0.01$. As expected, when both delays are zero, the equilibrium values of  $n_C^{A*}$ and  $n_D^{A*}$ are equal to $150$. However, when one delay is greater than the other one, the equilibrium point shifts, indicating a change in the system’s behavior.

\begin{figure}
    \centering
    \includegraphics[scale=0.22]{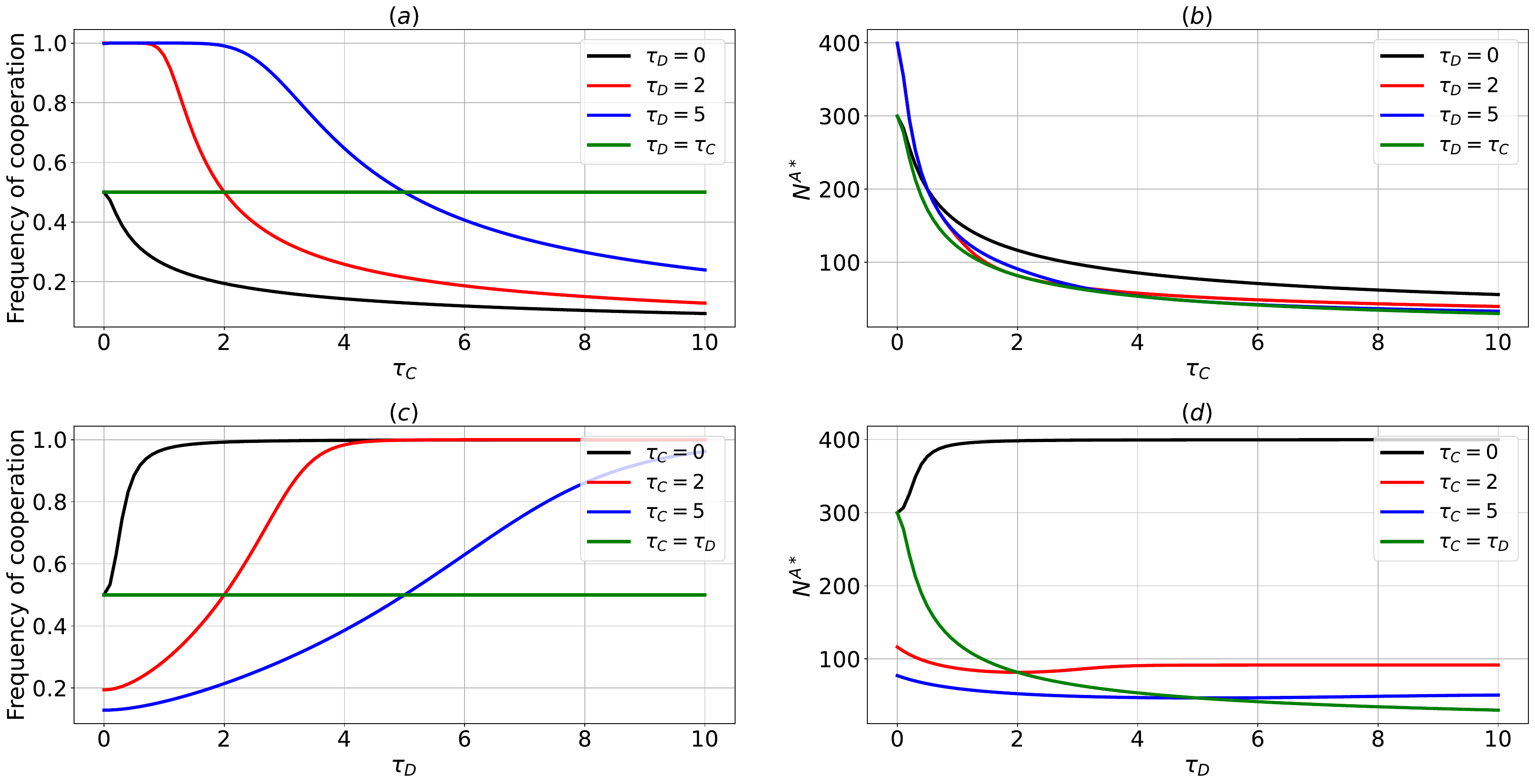}
    \caption{Frequencies of cooperation in the adult population and total adult population sizes in the stationary state as functions of time delays, numerical solutions of \eqref{deterministic2} with left-hand sizes equal to $0$. }
    \label{determindelay2}
\end{figure}

Fig. \ref{determindelay2} shows the effect of delays on the stable interior state ($x^*, N^{A*}$)
(stability verified through numerical solutions of Eq. (2)). Panels (a) and (c) show the impact of $\tau_C$ and $\tau_D$ on the stationary frequency of cooperation when the other delay is fixed, while the right panels show the dependence of the size of the adult population on delays. We observe that delays are not good for strategies, bigger the strategy delay, smaller its basin of attraction. A similar result was previously demonstrated in \cite{jamarek} for a system of delayed differential replicator equations and in \cite{fic} for the compartmental model. As noted earlier, for our payoff matrix, the stationary state is $0.5$, and this value remains unchanged when the delays for both strategies are increased simultaneously ($\tau_C = \tau_D$). It is possible for one strategy in the adult population to disappear entirely if the delay for that strategy significantly exceeds the delay for the other one. An interesting observation is that the same time delay can reduce the size of the adult population without affecting the stationary frequency.

In the next section, we incorporate stochasticity into our model to explore how delays influence population dynamics in the presence of stochastic fluctuations.

\section{Markov jump process of the compartment model}

Here we study a stochastic model corresponding to replicator dynamics discussed in the previous section. We follow closely Tao and Cressman \cite{taocressman}. In our construction, rates present in replicator equations become now intensities in a Markov jump process, a Markov chain in continuous time and a discrete space of states. 

The state of the Markov chain is described by the numbers of players of two strategies of both compartments, $\{n_i^j\}$, 
where $i \in \{C, D\}$ and $j \in \{A, K\}$. Intensities of jumps are given in Table \ref{table2}. 

\begin{table}[htbp]
\begin{tabular}{|l|l|}
\hline
transition& intensities \\ \hline
$n_i^A\to n_i^A+1$ & $\frac{n_i^K}{\tau_i}$  \\ \hline
$n_i^A\to n_i^A-1$ & $\beta^AN^An_i^A$\\ \hline
 $n_i^K\to n_i^K+1$& $n_i^A\pi_i$\\ \hline
 $n_i^K\to n_i^K-1$& $\frac{n_i^K}{\tau_i}+\beta^KN^Kn_i^K$\\\hline
\end{tabular}
\caption{Intensities of transitions between states of the Markov chain. The state of the chain at any time $t$ is given by $\{n_i^j\}$, where $i \in \{C, D\}$ and $j \in \{A, K\}$}
\label{table2}
\end{table}

Let $\Phi(n_C^A, n_D^A, n_C^K, n_D^K; t)$ be the joint probability distribution for the system to be at a given state at time $t$. One can write a standard Master equation, 

\begin{align}
    \frac{d \Phi(n_C^A,n_D^A,n_C^K,n_D^K;t)}{dt}&=\sum_{i= C,D}\sum_{j = A,K}(E_i^{+, j}-1)\beta^jN^jn_i^j\Phi(n_C^A,n_D^A,n_C^K,n_D^K;t)\nonumber\\&+\sum_{i =C,D}(E_i^{-,K}-1)n_i^A\pi_i \Phi(n_C^A,n_D^A,n_C^K,n_D^K;t) \nonumber\\
    &+\sum_{i = C,D}(E_i^{-,A}E_i^{+,K}-1)\frac{n_i^K}{\tau_i}\Phi(n_C^A,n_D^A,n_C^K,n_D^K;t),
    \label{masterequation}
\end{align}
where $E_i^{\pm,j}$ is the operator given by
\begin{align}
    E_i^{\pm,j}F(n_C^A,n_D^A,n_C^K,n_D^K)=
    \begin{cases}
       F(n_C^A \pm 1,n_D^A,n_C^K,n_D^K) \hspace{20pt}i=C,j=A \nonumber\\
       F(n_C^A,n_D^A \pm 1,n_C^K,n_D^K) \hspace{20pt}i=D,j=A \nonumber\\
       F(n_C^A,n_D^A,n_C^K \pm 1,n_D^K) \hspace{20pt}i=C,j=K \nonumber\\
       F(n_C^A,n_D^A,n_C^K,n_D^K \pm 1) \hspace{20pt}i=D,j=K
    \end{cases}
\end{align}
for any function $F$. It changes the population of strategists in adult or kindergarten by $\pm$ one individual. For example $E_C^{+, A}\{\beta^AN^An_C^A\Phi(n_C^A,n_D^A,n_C^K,n_D^K;t)\}= \beta^A(N^A+1)(n_C^A+1)\Phi(n_C^A+1,n_D^A,n_C^K,n_D^K;t)$.

One may then derive differential equations for the expected values and other moments of the number of players. However it is impossible to solve such equations. Therefore we will resort to stochastic simulations, we will implement a classic
Gillespie algorithm \cite{gillespie}.

Let us notice first that our Markov process has an absorbing state $(0,0,0,0)$. Such a situation is typical in many processes in biological and social models, the Moran process being the classic example. Our focus here is on the long-term behavior of the system under the assumption that the population is not extinct, we say that the system is in the quasi-stationary state. We will estimate expected values of population sizes and strategy frequencies in the quasi-stationary state.

We simulate $1000$ trajectories, with each trajectory consisting of $100,000$ Monte-Carlo steps. The expected value of $n_i^j$ is then computed as the average over the final $1000$ steps of all trajectories. Similarly, we calculate  $\langle x \rangle = \langle \frac{n_C^A}{N^A} \rangle$.

\begin{figure}
    \centering
    \begin{subfigure}{0.48\textwidth}
        \centering
        \includegraphics[width=\linewidth]{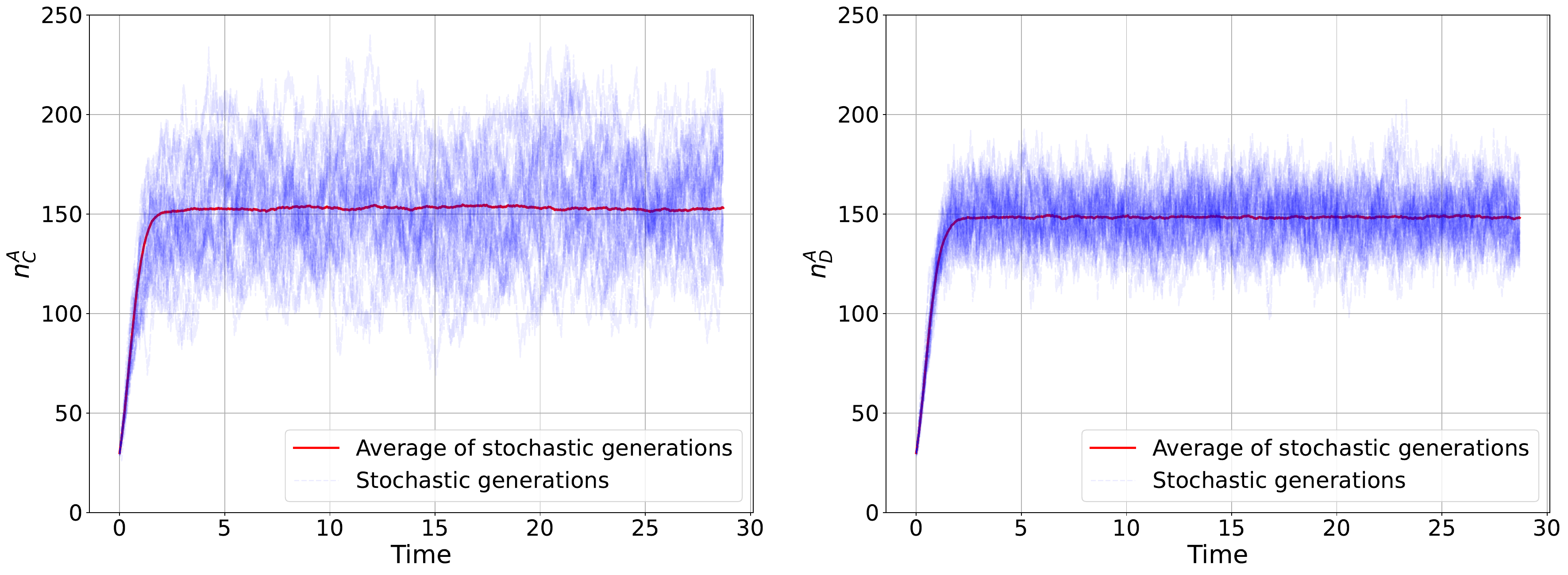}
        \caption{$\tau_C=\tau_D=0$}
        \label{fig:figure1}
    \end{subfigure}
    \hfill
    \begin{subfigure}{0.48\textwidth}
        \centering
        \includegraphics[width=\linewidth]{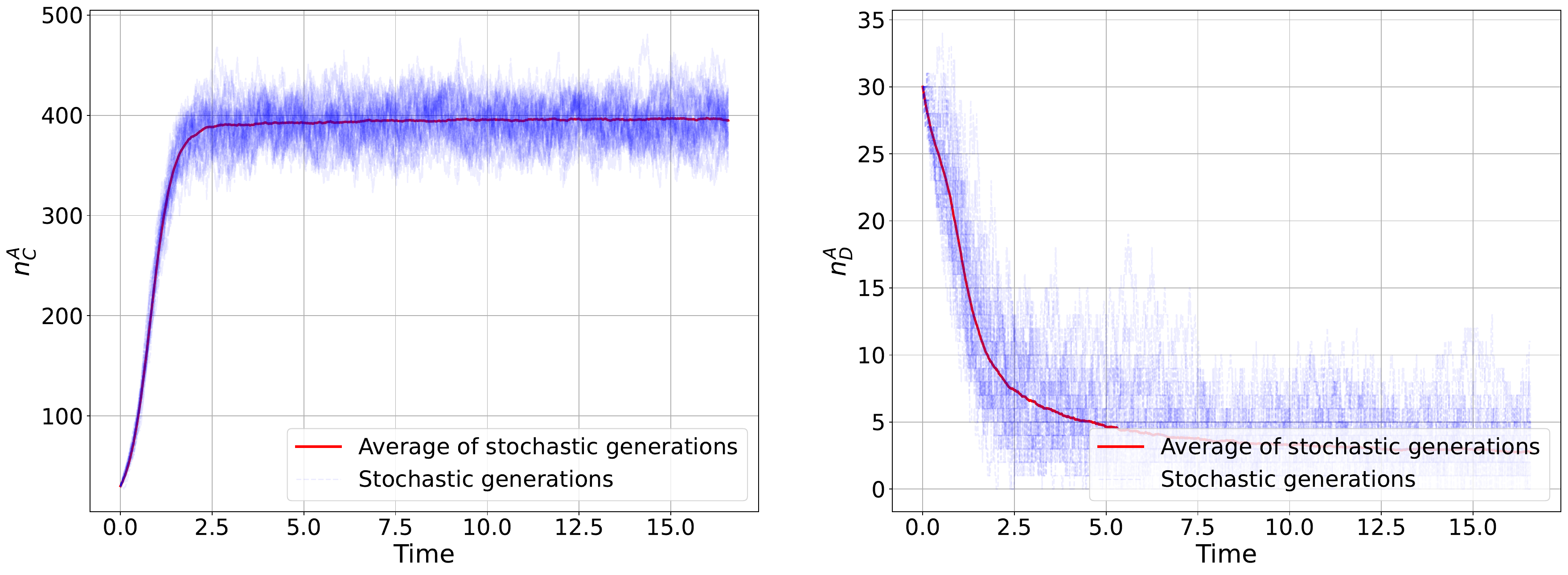}
        \caption{$\tau_C=0,\tau_D=2$}
        \label{fig:figure2}
    \end{subfigure}
    
    \vspace{0.5cm} 
    \begin{subfigure}{0.48\textwidth}
        \centering
        \includegraphics[width=\linewidth]{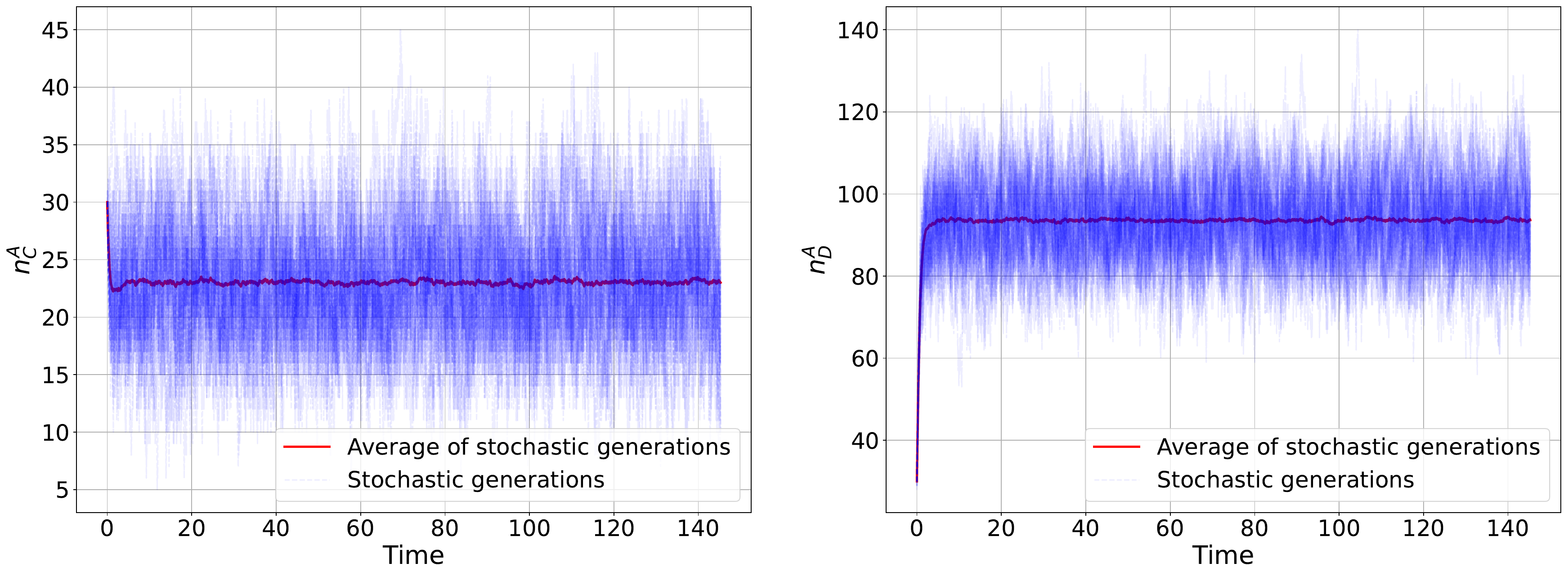}
        \caption{$\tau_C=2,\tau_D=0$}
        \label{fig:figure3}
    \end{subfigure}
    \hfill
    \begin{subfigure}{0.48\textwidth}
        \centering
        \includegraphics[width=\linewidth]{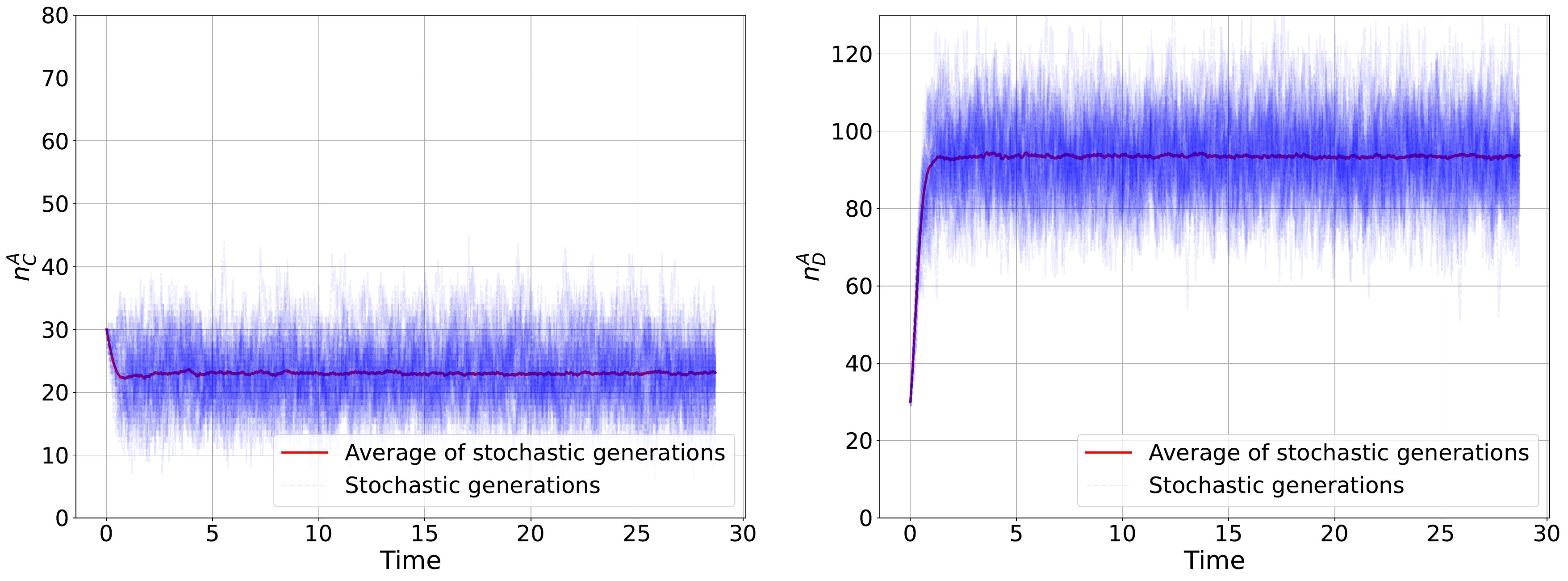}
        \caption{$\tau_C=2,\tau_D=2$}
        \label{fig:figure2}
    \end{subfigure}
    \caption{Single trajectories are represented by blue lines, their averages are shown in red, $\beta^A$ = $\beta^K$ = $0.01$ and the initial condition for each trajectory is set to $(30, 30, 0, 0)$}.
    \label{trajectories}
\end{figure}

Fig. \ref{trajectories} present stochastic trajectories for different delays, $\beta^A$ and $\beta^K$ are fixed at $0.01$ and the initial condition is set to $(30, 30, 0, 0)$, single trajectories are represented by blue lines, their averages are shown in red. All trajectories remain around the quasi-stationary state for an extended period of time. Since the only absorbing state of this process is $(0, 0, 0, 0)$, the system will eventually reach extinction. However starting from a big initial conditions and small $\beta^A$ and $\beta^K$, the extinction happens in a very long time. 

\begin{figure}
    \centering
    \includegraphics[scale=0.22]{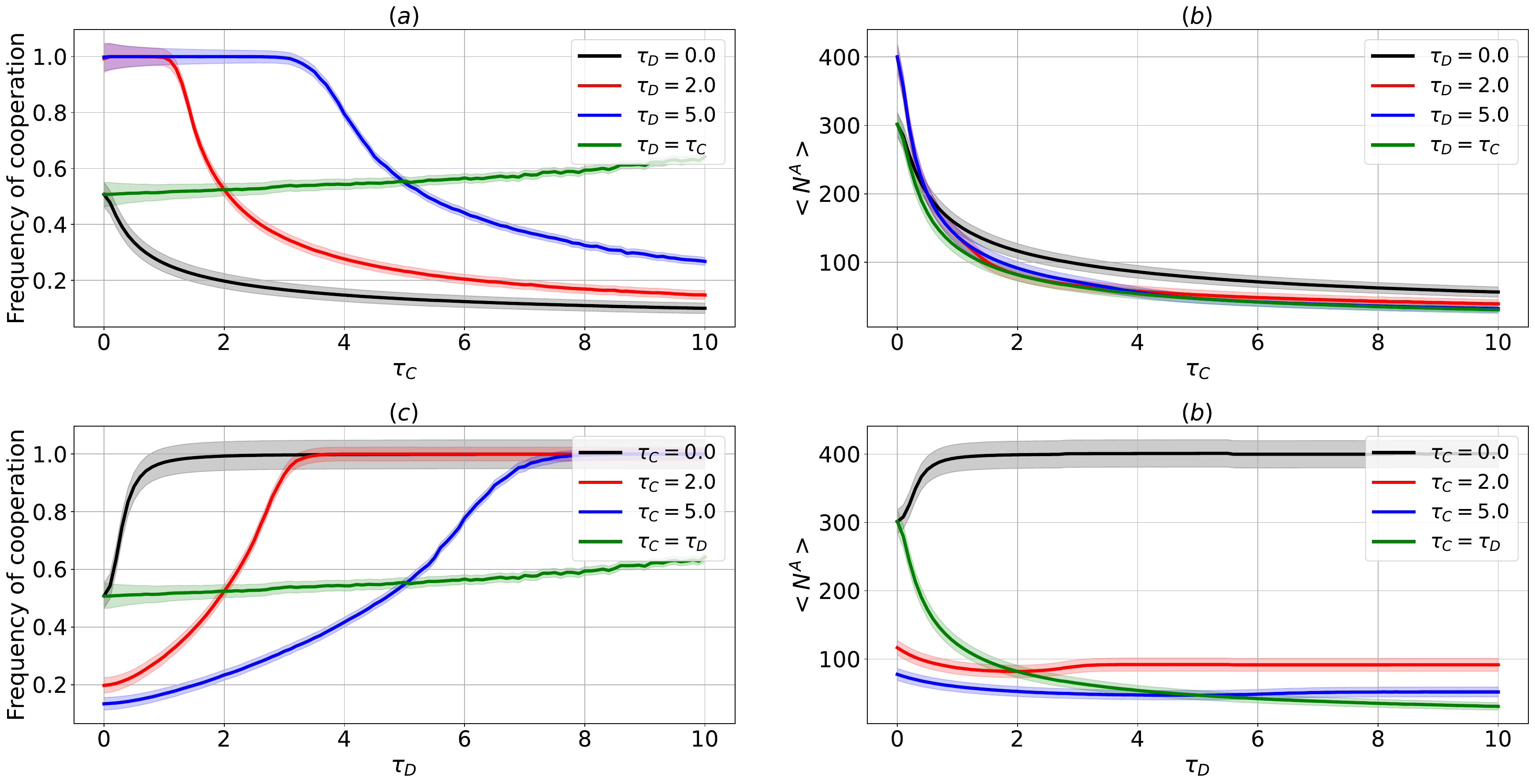}
    \caption{Expected values of the total adult population size and the frequency of cooperation $\langle x \rangle$ in the quasi-stationary state as functions of time delays. The shaded region around each line represents the standard deviation across all trajectories of the stochastic simulation.}
    \label{expectedvalues}
\end{figure}

Fig. \ref{expectedvalues} shows the dependence of 
the expected frequency of cooperators in the adult compartment and the size of the adult population in the quasi-stationary state. 

We observe fundamental differences between deterministic and stochastic evolution of our structured population.
In the stochastic model, Fig. \ref{expectedvalues}, panel (a), when the defector’s delay is fixed at $5$ and the cooperator’s delay varies, we observe that cooperation takes the whole adult population for cooperator delays below approximately $3$. In contrast, in the deterministic model, Fig. \ref{determindelay2}, panel (a), this threshold is lower, at around $2$. This suggests that in the stochastic dynamics, cooperation can persist in the population even at higher values of the delay parameter $\tau_C$ than in the deterministic dynamics. A similar pattern is seen when we compare panel (c) of Fig. \ref{expectedvalues} for the stochastic model and panel (c) of Fig. \ref{determindelay2} for the deterministic one. When the cooperator’s delay $\tau_C$ is fixed and the defector’s delay $\tau_D$ is varied, the stochastic model shows full cooperation in the adult population at smaller values of $\tau_D$ compared to the deterministic model.

\begin{figure}[htbp]
    \centering
    \includegraphics[scale=0.2]{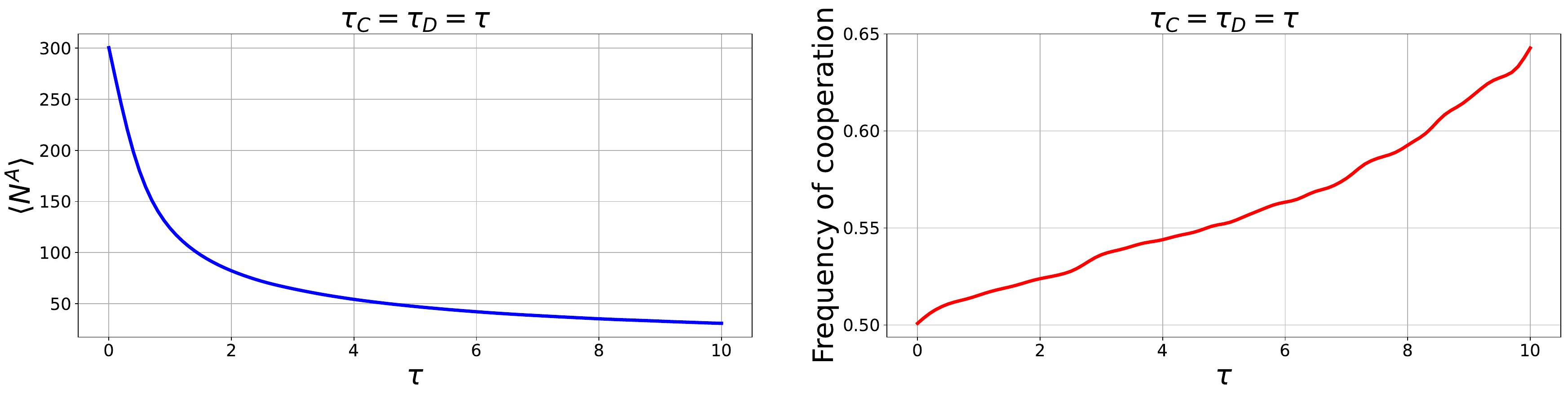}
    \caption{Expected values, $\langle N^A \rangle$ and  $\langle x \rangle$, in the quasi-stationary state as functions of delay equal for both strategies.}
    \label{fig:same delay1}
\end{figure}

Let us now compare both dynamics in the case of strategy-independent delays. We set $\tau_C = \tau_D = \tau$ and vary $\tau$. The results are shown in Fig. \ref{fig:same delay1}. When $\tau$ increases, the size of adult populations decreases. This outcome is expected, as individuals spend more time in juvenile compartments. However, we observe a novel behavior: the frequency of cooperators $\langle x \rangle$ in adult population increases with increasing delay. This observation contrasts with previous findings in \cite{jamarek,fic} where for strategy-independent delays there are no changes in frequencies of strategies is the stationary state of deterministic dynamics. Here strategy-independent delays promote the cooperative behavior in our stochastic model but not in the deterministic one.

\begin{figure}
    \centering
    \includegraphics[scale=0.3]{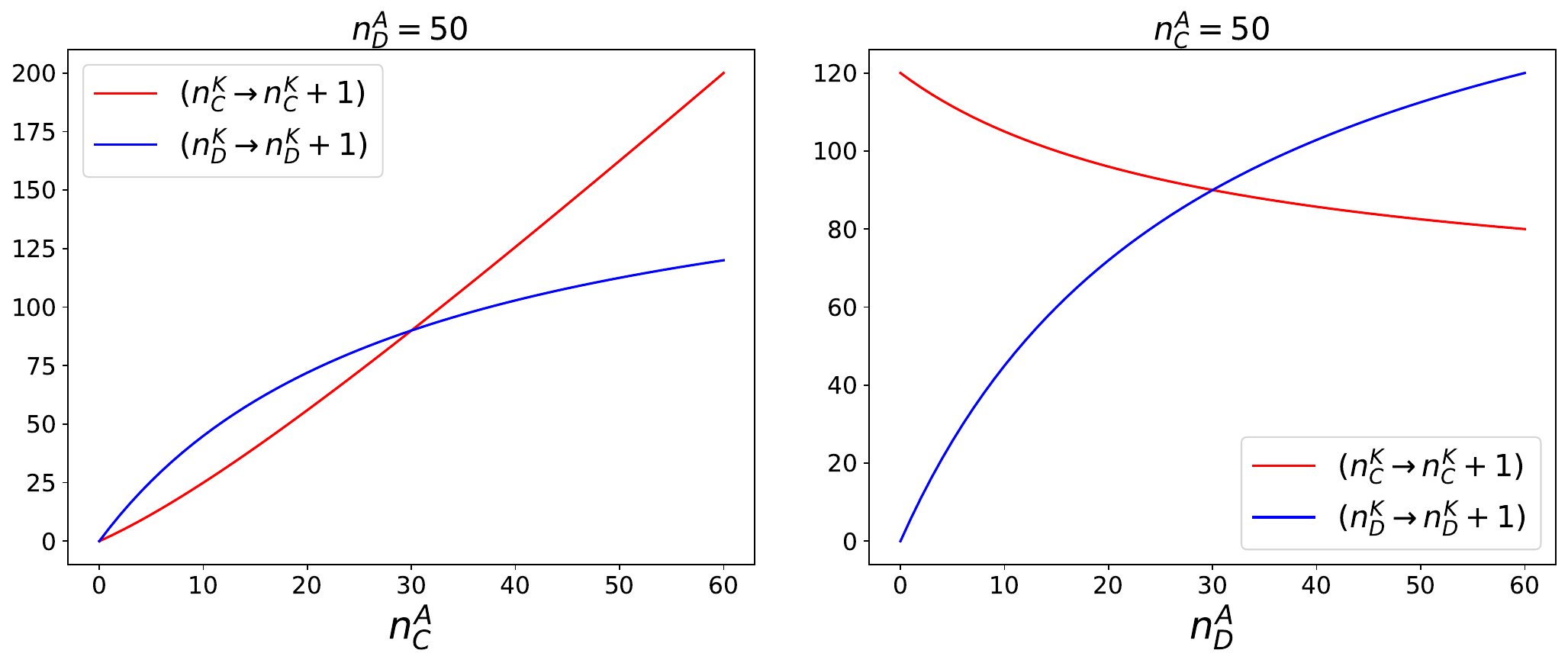}
    \caption{The intensity of a birth of a cooperator $n_C^A\frac{4n_C^A+2n_D^A}{n_C^A+n_D^A}$ and of a birth of a defector $n_D^A\frac{6n_C^A}{n_C^A+n_D^A}$.}
    \label{fig:explain}
\end{figure}

To understand why strategy-independent delays can promote cooperation, we analyze the intensity and probability of transitions in our Markov jump process. Let us assume initial conditions with equal number of cooperators and defectors in the adult population, $n_C^A = n_D^A = 30$. Because such initial conditions correspond to strategy frequencies in the mixed Nash equilibrium in our Snowdrift game, this is also a frequency stationary point of the deterministic structured dynamics, see Appendix A.

Let us recall that in the stochastic dynamics the intensity of the cooperator birth,
$n_C^K \to n_C^K + 1$, is given by 
$n_C^A \frac{4n_C^A + 2n_D^A}{n_C^A + n_D^A},$ 
while for the defectors birth it is $n_D^A \frac{6n_C^A}{n_C^A + n_D^A}$. At the initial condition both intensities are the same. However, in our stochastic process, only one event can occur at a time which can lead to an imbalance between the number of strategies. This imbalance influences intensities. Consider a simple example of having  additional cooperator or defector which happens with same probability $0.5$. It can be easily calculated that the average intensity in the next step of the birth of a cooperator is 91.51, whereas for the birth of a defector is slightly lower and equal to 91.47. This suggests that, on average, cooperators have a slight advantage. 

Fig. \ref{fig:explain} illustrates this behavior for scenarios where either the number of cooperators is fixed and the number of defectors varies, or vice versa. Consistent with the case of having one additional cooperator or defector, the results show that having more cooperators in the population is good for their growth. This bias toward cooperation suggests that, over time, the stochastic process will favor an increase in cooperators.

\begin{figure}[htbp]
   \centering
    \includegraphics[scale=0.22]{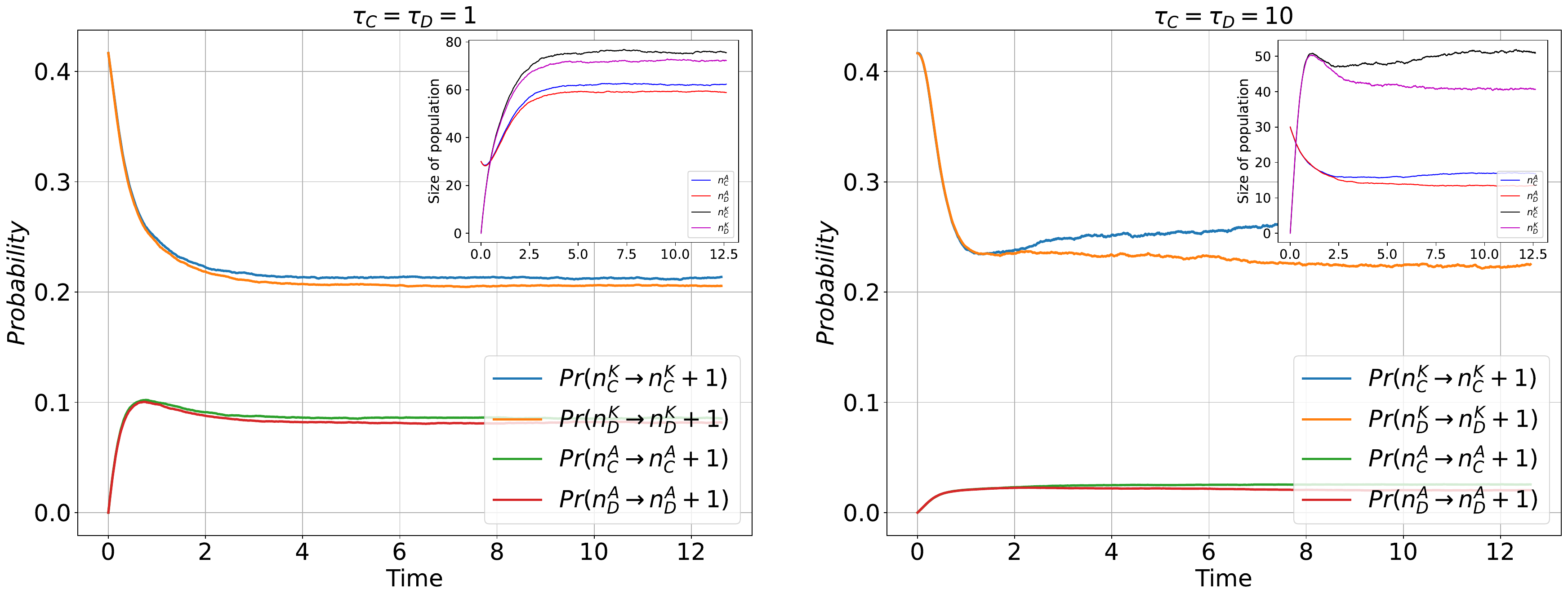}
    \caption{Time evolution of population sizes and probabilities of transitions for strategy-independent delays. The probability of a transition is the ratio of the intensity of that transition and the sum of all intensities at a current state of the system. Each curve is the average of 1,000 trajectories.}
    \label{fig:timeevaluation}
\end{figure}

When time delays are bigger, then juvenile spend more time in kindergartens and therefore the stationary size of the adult population is smaller, as seen in Fig. \ref{fig:same delay1}. One can see that for smaller initial conditions the above difference between intensities is bigger hence cooperators are more favored. In Fig. \ref{fig:timeevaluation}, we present the time evolution of frequencies and transition probabilities (ratios of a particular intensity and the sum of all intensities at a given state) for two different delay values: a small delay of 1 and a larger delay of 10. The figure illustrates that for a larger time delay, the difference in birth probabilities between cooperators and defectors is more significant compared to the smaller delay. This leads to a higher prevalence of cooperators when the time delay is bigger.

Below we examine the case where delays are not equal but are very close to each other. Previous studies have shown that when the delay associated with a given strategy is larger than the delay of the other one, then it tends to disadvantage that strategy.

\begin{figure}[htbp]
    \centering
    \includegraphics[scale=0.4]{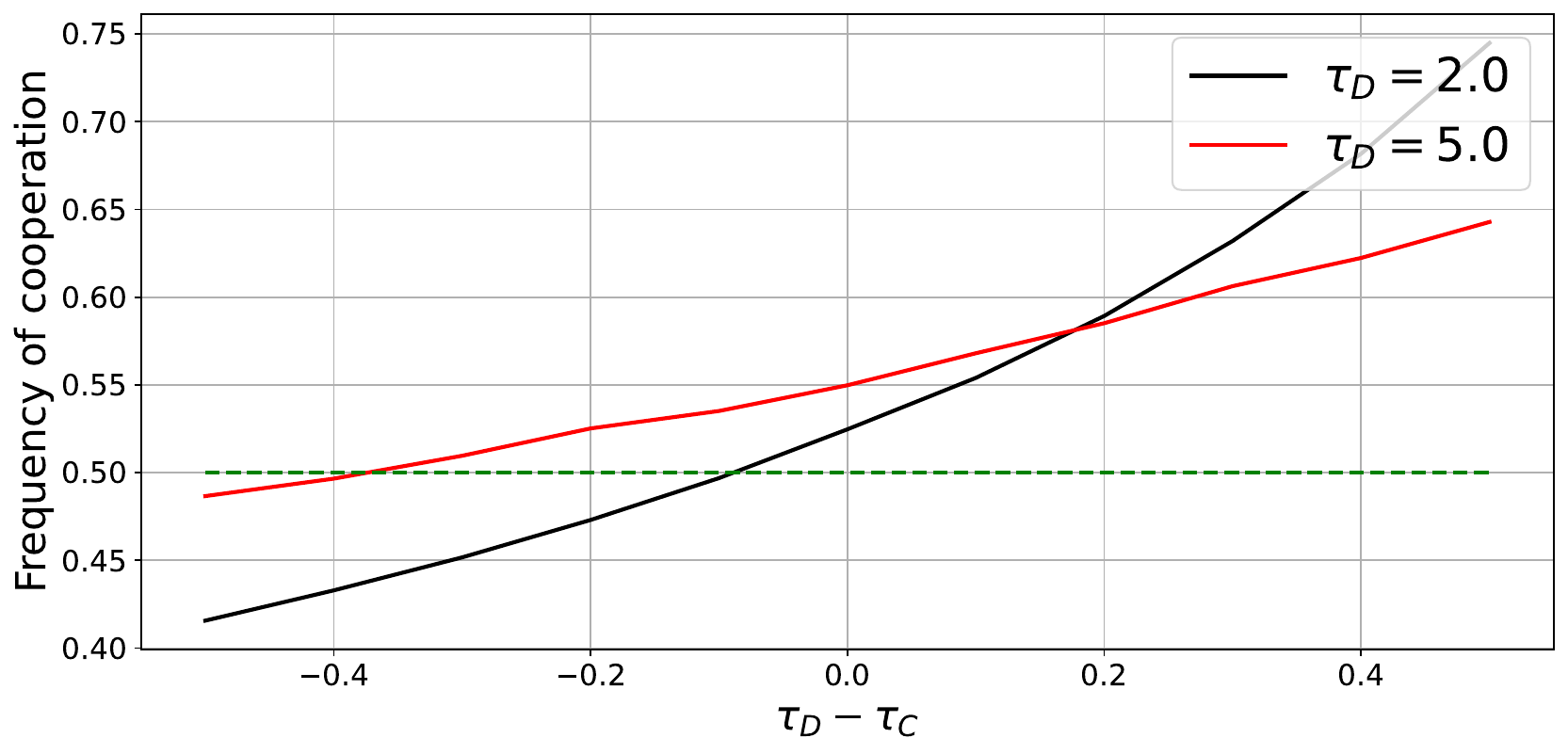}
    \caption{Cooperation frequency in the 
    quasi-stationary state as a function time delays difference.}
    \label{fig:same delay2}
\end{figure}

However, in Fig. \ref{fig:same delay2}, we observe that for a game with an interior stationary state at $0.5$ (in the absence of delays), even a slightly larger delay for $\tau_C$ compared to $\tau_D$  promotes cooperation. In such cases, the quasi-stationary frequency of cooperators exceeds $0.5$. Of course, this effect holds only when $\tau_C$ is slightly larger than $\tau_D$. When the difference becomes significant, and $\tau_C$ is sufficiently large, similar to what is seen in the deterministic, it negatively impacts cooperation, reducing the cooperator frequency below $0.5$. The value of $\tau_C$ (where $\tau_C > \tau_D$) at which cooperation is still favored depends on the value of $\tau_D$. For instance, in Fig.\ref{fig:same delay2}, when $\tau_D = 2$, cooperation is favored up to $\tau_C = 2.1$. However, when $\tau_D = 5$, cooperation can be favored up to $\tau_C = 5.35$. 

\begin{figure}[htbp]
    \centering
    \includegraphics[scale=0.55]{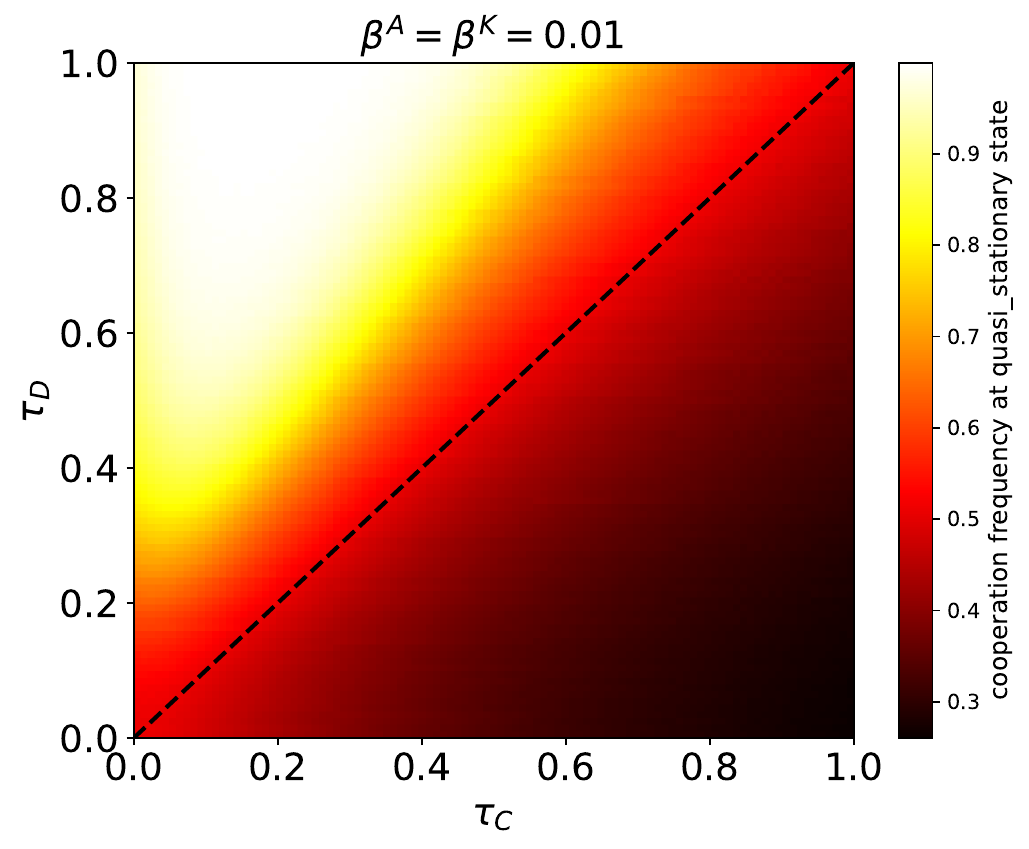}
    \caption{Effect of $\tau_C$ and $\tau_D$ on the frequency of cooperation in the adult population in the quasi-stationary state.}
    \label{fig:imshow}
\end{figure}

Fig. \ref{fig:imshow} illustrates effects of both delays on the frequency of cooperators in the quasi-stationary state. Clearly, for a fixed $\tau_D$, increasing $\tau_C$ leads to a decrease in the level of cooperation. Conversely, for a fixed $\tau_C$, increasing $\tau_D$ results in the opposite trend, with cooperation levels rising. 

\section{Discussion}\label{sec13}

We considered a Snowdrift game which has an asymptotically stable interior stationary state in the replicator dynamics.
We combined ideas and constructions of \cite{taocressman} and \cite{fic} and presented a structured replicator dynamics 
of finite populations and the corresponding birth-death Markov jump process with two compartments: adults and juveniles. 
In the stochastic dynamics, inverses of rates of leaving juvenile compartments may be interpreted as averages of random exponentially distributed time delays. 
Our main result is that for a small difference of inverse rates (average time delays) of both strategies, the frequency of cooperators 
in the quasi-stationary state of the stochastic dynamics is bigger than that in the corresponding stationary state 
of the deterministic structured replicator dynamics; for equal average delays, such stationary state is equal to the critical point 
of the original replicator dynamics for the Snowdrift game. In short, time delays favor cooperation in the presence of intrinsic stochastic fluctuations.

Somewhat analogous behavior has been observed recently in random walks with asymmetric time delays. It was observed there 
that we may reverse effects of time delays by a symmetric transformation of fitness functions and then the time delay 
of a given strategy increases its frequency in the stationary distribution \cite{randomwalksdelays}. 

It is worth to mention here that a logistic suppression was directly incorporated in replicator dynamics in \cite{argbroom1,argbroom2,argbroom3}. 
It is an interesting problem to extend the authors models by adding time delays and intrinsic noise. 

In general it is important to study the joint effects of time delays and stochasticity. 
Some results were presented in \cite{jawes,threeplayers,randomwalksdelays,javbroom}.

\section*{Acknowledgements}
This project has received funding from the European Union’s Horizon 2020 research and innovation program 
under the Marie Sk\l odowska-Curie grant agreement No 955708.

\section*{Appendix A}\label{secA1}
\renewcommand{\thesection}{A} 
\setcounter{section}{1}
As mentioned in the main text, the deterministic equations differ when either one or both delays are zero. If the delay for a particular strategy is zero, it implies that new juveniles, born in proportion to the payoff, are immediately added to the adult population. Below, we present the equations for the various delay scenarios.

\begin{itemize}
    \item $\tau_C=\tau_D=0$:
    \begin{align}
    \begin{cases}
    \frac{dn_C^A}{dt}=n_C^A\pi_C-\beta^AN^An_C^A\\
    \frac{dn_D^A}{dt}=n_D^A\pi_D-\beta^AN^An_D^A      
    \end{cases}
    \end{align}

    \begin{align}
    \begin{cases}
        \frac{dx}{dt}=x(1-x)(\pi_C-\pi_D)\\
        \frac{dN^A}{dt}=N^A(\bar{\pi}-\beta^AN^A)
    \end{cases}
    \end{align}
    \item $\tau_C=0, \tau_D>0$:
\begin{align}
\begin{cases}
    \frac{dn_C^A}{dt}=n_C^A\pi_C-\beta^AN^An_c^A\\
    \frac{dn_D^A}{dt}=\frac{n_D^K}{\tau_D}-\beta^AN^An_D^A\\
    \frac{dn_D^K}{dt}=n_D^A\pi_D-\beta^K{(n_D^K)}^2-\frac{n_D^K}{\tau_D}
    \end{cases}
    \end{align}
\begin{align}
\begin{cases}
    \frac{dx}{dt}=x\big((1-x)\pi_c-\frac{N^K}{\tau_D}\big)\\
    \frac{dN^A}{dt}=N^A\big(x\pi_c-\beta^AN^A\big)+\frac{N^K}{\tau_D}\\
    \frac{dN^K}{dt}=N^A(1-x)\pi_D-\frac{N^K}{\tau_D}-\beta^K{N^K}^2
\end{cases}
\end{align}
    \item $\tau_C>0, \tau_D=0$:
        \begin{align}
        \begin{cases}
         \frac{dn_D^A}{dt}=n_D^A\pi_D-\beta^AN^An_D^A\\
         \frac{dn_C^A}{dt}=\frac{n_C^K}{\tau_C}-\beta^AN^An_C^A\\
        \frac{dn_C^K}{dt}=n_C^A\pi_C-\beta^K{(n_C^K)}^2-\frac{n_C^K}{\tau_C}
            \end{cases}
            \end{align}
    \begin{align}
    \begin{cases}
        \frac{dx}{dt}={(1-x)}^2\pi_D-\frac{N^Kx}{\tau_C}\\
        \frac{dN^A}{dt}=N^A\big((1-x)\pi_D-\beta^AN^A\big)+\frac{N^K}{\tau_C}\\
        \frac{dN^K}{dt}=N^Ax\pi_C-\frac{N^K}{\tau_C}-\beta^K{N^K}^2
        \end{cases}
        \end{align}
    \item $\tau_C=\tau_D=\tau$:
    \begin{align}
    \begin{cases}
    \frac{dn_C^A}{dt}=\frac{n_C^K}{\tau}-\beta^AN^An_C^A\\
    \frac{dn_D^A}{dt}=\frac{n_D^K}{\tau}-\beta^AN^An_D^A\\
    \frac{dn_C^K}{dt}=n_C^A\pi_C-\beta^KN^Kn_C^K-\frac{n_C^K}{\tau}\\
    \frac{dn_D^K}{dt}=n_D^A\pi_C-\beta^KN^Kn_D^K-\frac{n_D^K}{\tau}
    \label{determinestic}
    \end{cases}
    \end{align}
\begin{align}
\begin{cases}
    \frac{dx}{dt}=\frac{N^K}{N^A}\big(\frac{y(1-x)-x(1-y)}{\tau}\big)\\
    \frac{dy}{dt}=\frac{N^A}{N^K}\big(x\pi_C-y\bar{\pi}\big)\\
    \frac{dN^A}{dt}=\frac{N^K}{\tau}-\beta^A(N^A)^2\\
    \frac{dN^K}{dt}=N^A\bar{\pi}-\frac{N^K}{\tau}-\beta^K{N^K}^2
    \end{cases}
    \label{equal}
    \end{align}
\end{itemize}
The key observation in our paper is that the qualitative behavior is different for deterministic and stochastic dynamics when delays for both strategies are equal. 
We solve the system of equations \eqref{equal} for the stationary state by setting all derivatives to zero. From the first equation we get $x^* = y^*$. Substituting this into the second one we obtain $\frac{N^A}{N^K}x(1-x)(\pi_C-\pi_D)=0$. Hence the stationary interior state is given by $\pi_C = \pi_D$
as in the classical replicator dynamics without time delays.

We conclude that strategy-independent delays do not shift stationary states of frequencies in the deterministic dynamics.



\noindent


\bibliographystyle{abbrv}

\end{document}